\documentclass{aa}
\usepackage{graphics}
\usepackage{amssymb}

\newcommand{\fig}[1]{Fig. \ref{#1}}
\newcommand \msun   {\mbox{ M$_\odot$}}
\newcommand \etal   {et al. }
\newcommand \degree  {\mbox{$^\circ$}}
\newcommand{\tab}[1]{Table \ref{#1}}
\begin{document}

   \thesaurus{06     
              (03.11.1;  
               16.06.1;  
               19.06.1;  
               19.37.1;  
               19.53.1;  
               19.63.1)} 
   \title{Interacting star clusters in the LMC \thanks{Based on observations 
made at the European Southern Observatory, La Silla, Chile}}

  \subtitle{Overmerging problem solved by cluster group formation}

   \author{St\'ephane Leon \inst{1,2}, Gilles Bergond \inst{2,3}, and Antonella 
Vallenari \inst{4}}

   \offprints{S. Leon ({\tt sleon@mesunb.obspm.fr})}

   \institute{  DEMIRM, Observatoire de Paris, 
		61 Av. de l'Observatoire,
		F-75014 Paris, France
		\and
		CAI-MAMA,
		Observatoire de Paris,
		61, Av. de l'Observatoire,
		F-75014 Paris, France
		\and
		DASGAL, Observatoire de Paris-Meudon,
		5, Place J. Janssen
		F-92195 Meudon, France
		\and
		Astronomical Observatory of Padova, 
		Viccolo dell'Osservatorio 5, 
		I-35122 Padova, Italy 
	}

   \date{Received XX; accepted XX}

   \maketitle

   \begin{abstract}
   We present the tidal tail distributions of a sample 
of candidate binary clusters located in the bar of the 
Large Magellanic Cloud (LMC). One isolated cluster, SL~268, 
is presented in order to study the effect of the LMC tidal 
field. All the candidate binary clusters show tidal tails, 
confirming that the pairs are formed by physically linked 
objects. The stellar mass in the tails covers a large range,
from $1.8\times 10^3$ to $3\times 10^4$\msun. We derive a 
total mass estimate for SL 268 and SL 356. At large radii, 
the projected  density profiles of SL~268 and SL~356 fall 
off as $r^{-\gamma}$, with $\gamma= 2.27$ and $\gamma=3.44$,
respectively. Out of 4 pairs or multiple systems, 2 are 
older than the theoretical survival time of binary clusters 
(going from a few $10^6$ yr to $10^8$ yr). A pair shows too 
large age difference between the  components to be consistent 
with classical theoretical models of binary cluster formation 
(Fujimoto \& Kumai 1997). We refer to this as the 
``overmerging'' problem.A different scenario is proposed: 
the formation  proceeds in large molecular complexes giving 
birth to groups of clusters over a few $10^7$ yr. In these 
groups the expected cluster encounter rate is larger, and 
tidal capture has higher probability. Cluster pairs are not 
born together through the splitting of the parent cloud,
but formed later by tidal capture. For 3 pairs, we 
tentatively identify the star cluster group (SCG) 
memberships. The SCG formation, through the recent cluster 
starburst triggered by the LMC-SMC encounter, in contrast 
with the quiescent open cluster formation in the Milky Way 
can be an explanation to the paucity of binary clusters 
observed in our Galaxy.
 
               
   \end{abstract}

%

\section{Introduction}

The LMC posses a large population of candidate binary clusters, having no 
counterpart in the Milky
Way where
only a few open clusters
are known to be binary, as h\&$\chi$ Persei (Subramiam \etal 1995).
 In the LMC the existence of binary 
clusters was disclosed 
a decade ago (Bhatia \&
Hatzidimitriou 1988; Bhatia \etal 1991).
Using statistical arguments,
Bhatia \& Hatzidimitriou (1988) claimed that  a considerable fraction of
the suspected binary clusters 
 must be physically 
linked.
 
However, on theoretical ground, their physical status and their
properties (formation process, survival time\ldots) remain unclear.
Oliveira \etal (1998) have performed $N$-body simulations of star cluster
encounters explaining the presence of some morphological effects found
in LMC cluster pairs as expanded halo, isophotal deformation and isophotal
twisting.
  Fujimoto \& Kumai (1997)
have explained the formation of pairs in terms of oblique
cloud-cloud collisions. As a result  of these collisions, the clouds
are split in two parts, forming the cluster pair. 
As a consequence, the two clusters have the same age.
The lifetime of a proto-cluster is unknown, but is is estimated to be
of $10^8$yr at maximum. This is the expected age difference between
the two components of cluster pairs.
Clearly,
this model runs into problems to account for  the presence
of non-coeval pairs of clusters as observed in the LMC (see Vallenari \etal 
1998). It  has been  proposed (Grondin 
\etal 1992; Vallenari 
\etal 1998) that the recent cluster  burst formation some $10^8$ years ago could 
have been 
triggered by the interaction between the SMC and the LMC. Recent works 
on Galactic clusters (Grillmair \etal 1995 (hereafter G95); Leon \etal
1998; Bergond \etal 1998) have shown that the  globular and open 
clusters disclose huge tidal tail extension due to
their interactions with the strong gravitational tidal field.
 In this framework, the 
study of the 
isolated and binary clusters in the LMC is interesting as probe 
of the star cluster 
 shaping by the
LMC tidal field. Moreover LMC star clusters are found to have higher ellipticity
 than their galactic 
counterparts: Goodwin (1997) has
suggested that tidal field of the parent galaxy is the dominant
factor  in determining ellipticity
of globular cluster. However high ellipticity can as well be the result
of a different process:
  Sugimoto \& Makino (1989) suggested that some of most elliptical
objects could be the 
rotating remnants of cluster mergers.
 They propose a 
  scenario
where some of young clusters  born in pairs from the proto-cluster gas, will  
eventually merge or be disrupted in a few $10^7$
years. Stellar halo 
 truncation by the
tidal interactions and dynamical friction  is decreasing progressively the 
ellipticity of the 
merger clusters. This paper is devoted to discuss the properties
of a sample of suspected binary clusters and the 
 implications on the formation scenario.
                                                                                          
The observations and the data reduction are presented in Section 2 and Section 3
respectively.  
The properties of individual cluster pairs are
discussed in Section 4 and, finally the conclusions are drawn in
Section 5.

\section{Observations and presentation of the cluster sample}

The observations have been made in La Silla on December 16th-17th 1995, using
the NTT ESO telescope equipped with EMMI and the ESO CCD \#36 (0.27$\arcsec/$pixel). More
details about the 
observations can be found in Vallenari \etal (1998) for all the clusters, except 
for NGC 1850.
NGC~1850 has been observed with the ESO 2.2m telescope in La Silla 
(0.46$\arcsec/$pixel). Observations are described
in Vallenari et al. (1994).
The stellar photometry has been performed 
using Daophot\,II and a conversion to the standard Johnson $V$ and $I$ has been 
done from 
calibration stars
in the Landolt list. On \tab{table_sample} we present the cluster sample located 
in the west side
part of the LMC: 6 are components of suspected
 binary systems (SL 349-SL 353, SL 356-SL 357, 
SL 385-SL 387).
NGC 1850 can be considered as a binary or even 
triple system 
(Fischer \etal 1993; Vallenari et al. 1994). 
SL 268 is an isolated object
 included  to have 
a control
cluster relative to the LMC gravitational tidal field.\\

\begin{table}
\begin{tabular}{lcccccc}
\hline
\hline
Cluster	& X  & Y & Age & Diam.  & Sepa. \\
	& (\degree)  &  (\degree) & (Log(yr)) & (pc) & (pc) \\
\hline
SL 268   & 1.04 & -0.11& 8.65  & 12 & \\
NGC 1850  & 1.11	& 0.70 & 7.95   & 12  & \\
SL 349	 & 0.62	& 0.40 &  8.70 & 12 & 19\\
SL 353   & 0.66 & 0.38 & 8.70  & 12 &  19\\
SL 356	& 0.32	& 0.14 & 7.85  & 15 & 39 \\
SL 357 			& 0.34 & 0.14 &  8.78 & 9  & 39 \\
SL 385			& 0.37 & 0.16 & 8.18 & 10 & 10 \\
SL 387			& 0.44 & 0.14 & 8.70 & 10 & 10 \\
\hline
\end{tabular}

\caption{Observed cluster properties: X and Y are the positions inside the LMC 
from
Bica \etal (1996). The age is from Vallenari \etal (1998), except for NGC 1850 
which is
from Fischer \etal (1993). The diameter is from Bica \etal (1996) and the 
separation is from 
Bhatia \etal (1991).}
\label{table_sample}
\end{table}

\section{Data reduction}

\subsection{Tidal tail extension}

\begin{figure}
\resizebox{8.5cm}{8cm}{\includegraphics{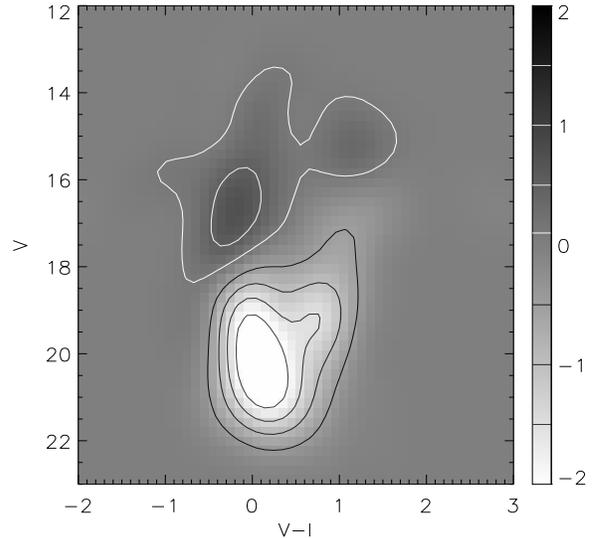}}
\caption{Function s-n towards NGC 1850. A high s-n value in the color-magnitude space 
indicates
a reliable separation for the underlying stars relative to the field stars.}
\label{fig_cmd_sn_n1850}
\end{figure}

From the Color-Magnitude Diagram (CMD) contrast between the cluster and the 
field, we construct
a signal/noise function s-n in the Color-Magnitude Space (CMS) for 
each pair: details of the method
 are given in G95. In \fig{fig_cmd_sn_n1850} the function s-n 
is presented for
NGC 1850: its main sequence termination point, because of its young age, is 
about $V\sim 16.5$
and the evolved star region is clearly seen with a good S/N. The low values of 
the function 
represent  the main sequence stars of the  LMC field population.

SL 349 and SL 353 are
not distinguishable on the basis of
 their CMD, due to their similar age. We will treat 
both clusters as
a whole  to derive the tidal extension of the pair. SL 268 and SL 356 present 
the higher 
contrast 
relative to the field stars.

\begin{figure}
\resizebox{\hsize 
}{!}{\includegraphics{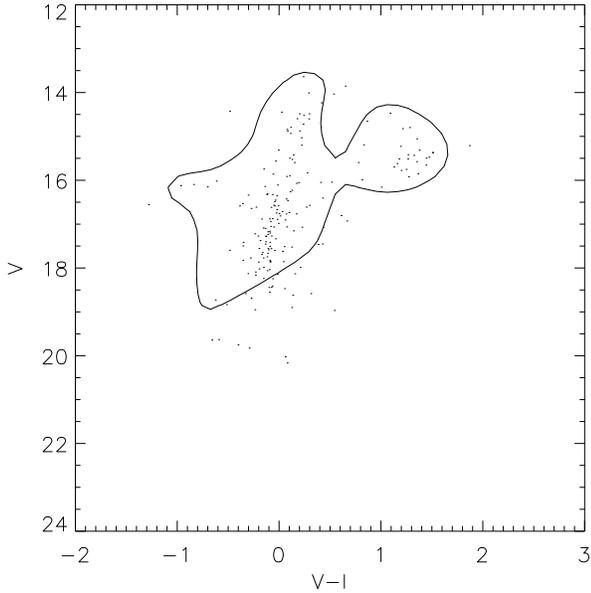}}
\caption{CMD of the stars towards NGC 1850 and the contour which limits the area 
with a high
S/N to select the stars.}
\label{fig_n1850_cmd_select}
\end{figure}

\begin{figure*}
\resizebox{\hsize }{!}{\includegraphics{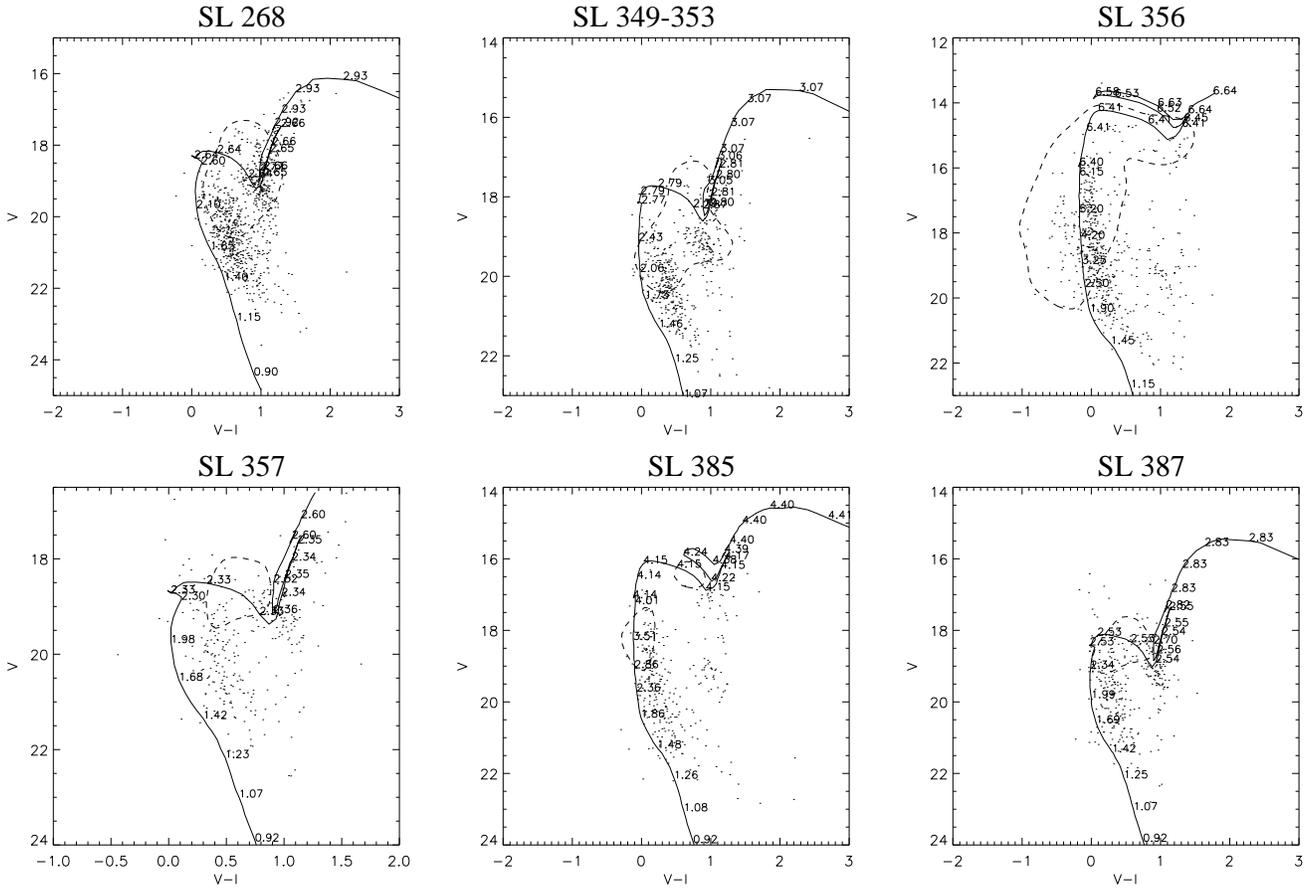}}
\caption{CMDs of the stars towards the clusters. The dashed line is the mask to 
select the stars. As we already said, the similarity of the two CMDs in the case 
of SL 349 and  SL 353 makes it impossible to find 
a reliable area in the
CMS to separate the two clusters. Isochrones of the appropriate age  
 are plotted (see  Table~1 and for more details
 Vallenari \etal 1998). Stellar masses along the isochrones (in\msun) are also 
indicated.}
\label{fig_cmd_select_all_iso}
\end{figure*}

 In all the cases the highest S/N is for the upper 
part of the main 
sequence and for the evolved stars. From this s-n function, we select a 
high-S/N region in the CMS
to filter the star catalog, as shown for NGC 1850 in
\fig{fig_n1850_cmd_select}. In \fig{fig_cmd_select_all_iso}  
we present  the areas in the CMS, 
retained to select
the stars for the whole sample of clusters.
 
\begin{figure*}
\resizebox{!}{23cm}{\includegraphics{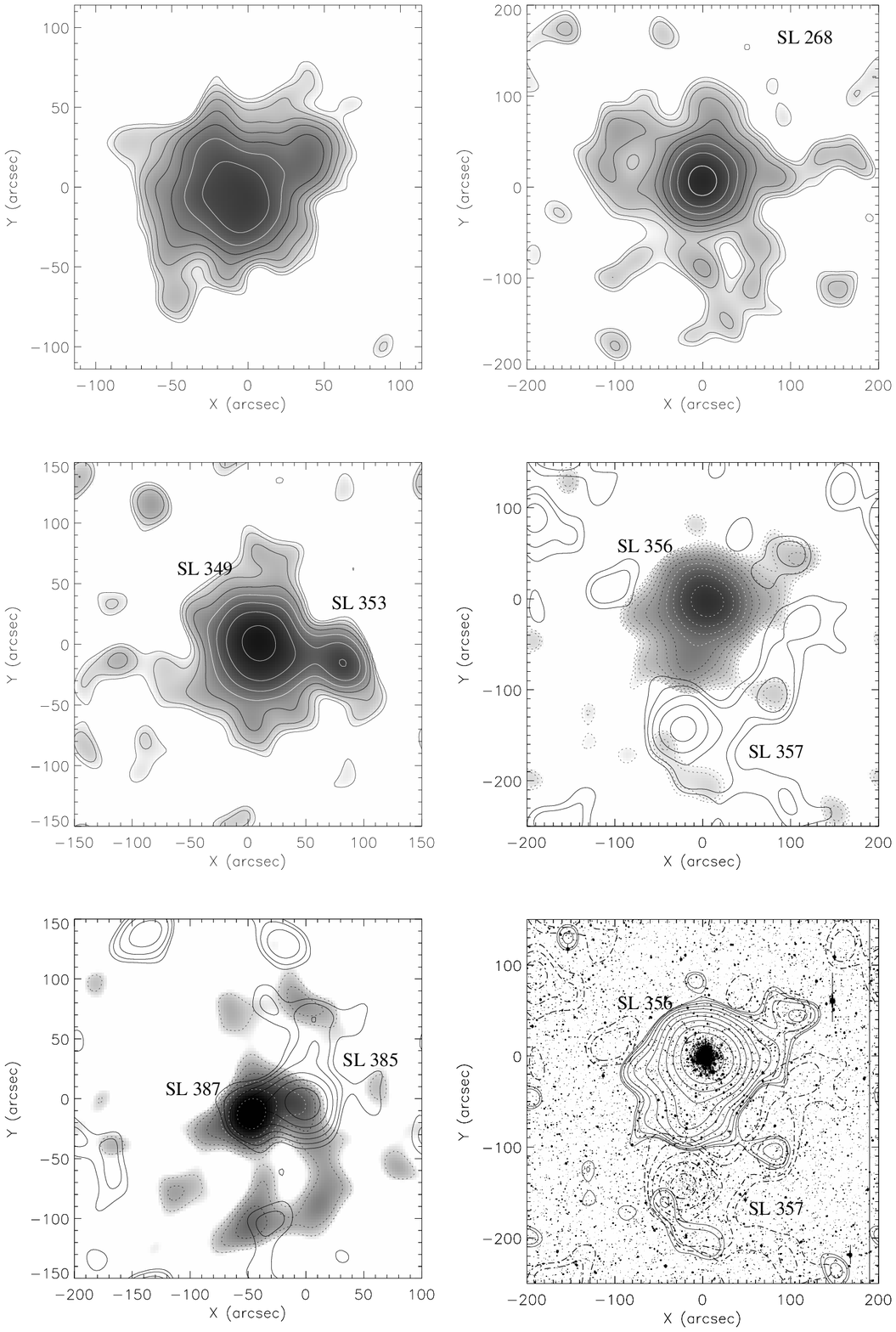}}
\caption{Surface density (Log) of the star clusters. For SL 356-357 and 
SL 385-387 contours stand for one of the components. In addition,  we present in the lower-right panel
 the optical image of the pair SL 356-357 overlaid by the same contours.}
\label{fig_tail_cluster}
\end{figure*}

The threshold is chosen as a compromise between  the highest S/N and 
star-counts sufficiently
high to prevent poissonian noise fluctuation which would decrease the final 
spatial resolution after the wavelet reconstruction: the higher is the S/N 
threshold the 
better is the separation between the stars field and the clusters and between 
the clusters 
themselves with the constrain to be above $3\sigma$ poissonian star-counts noise.
On the CMD-selected star counts map  we fit a background map 
following G95 by masking the cluster (1 to 2 $r_c$) and using a 
blanking value inside,
equal to the mean between 1.5 and 2.5 $r_c$ to get a smooth background: we fit
a low-order bivariate polynamial surface, mainly first or second order surface 
to avoid to erase some local variation by higher order polynomials. 
We subtract this background from the CMD-selected map. 
On that final density map  a Wavelet Transform (WT) is performed, 
using the so-called 
``\`a trous'' algorithm (see Bijaoui 1991) which 
decomposes a  2D array in different planes, each one being a representation of a 
particular scale.
A more complete discussion of the use of WT can be found in Leon \etal (1998). 
We retain
only the planes 3 to 7, removing the highest planes which represent the low 
scale details. Different
test on the final map have shown (Leon \etal 1998)  that this resolution can be 
used  at a 
high confidence level. Typically the map resolution in the cluster region is 
about 15\arcsec.
On \fig{fig_tail_cluster}  the tidal tail density distribution is shown.

\subsection{Tidal tail mass}

\begin{figure}
\resizebox{\hsize }{!}{\includegraphics{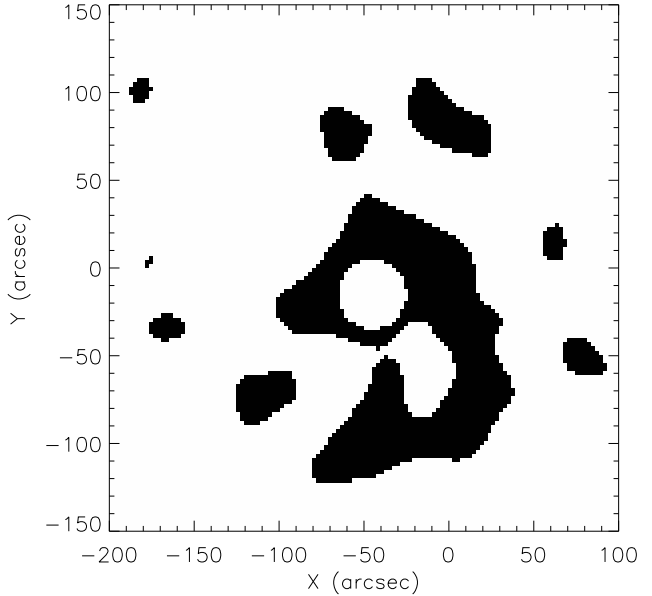}}
\caption{Example of mask on the tail of SL 387 used to count the stars selected 
from
the CMD mask. We mask as well the cluster using a radius from Bica \etal 
(1996).}
\label{fig_masque_sl387}
\end{figure}

We perform  star-counts in the tail by masking  the cluster itself, 
first using the radius
of the cluster $r_c$ given by Bica \etal (1996) and shown in Table~1,
and second, defining a radius $2 \times r_c$.
Then we 
mask the
star-count from the density above a threshold, which is chosen as the lowest 
level on
the \fig{fig_tail_cluster}. An example of mask, for SL 387, is shown on 
\fig{fig_masque_sl387}.

To derive the stellar mass released in the tail we estimate the ratio of stars 
in 
the CMD mask used to select the stars relatively to the total number of stars.
The stellar mass range in the clusters
 are derived from the isochrone fitting of the cluster CMDs  made by
Vallenari \etal (1998).
 In 
\fig{fig_cmd_select_all_iso} 
the CMDs and the appropriate isochrones are shown for each cluster,
together with the involved stellar masses.
We define for each object $M_{min}$ and $M_{max}$ as the minimum and maximum  
stellar masses
respectively,
 in selected mask
of the CMD. 
 We make use of a Salpeter
law for the initial mass function (IMF): 

\begin{equation}
\Phi(m) =K m^{-\alpha} \mbox{  , $m_{low} < m < m_{up}$}
\end{equation}
where  $\alpha$  is the  slope,
 $m_{low}=0.2$ and $m_{up}$ are the lowest and highest stellar
masses respectively, in the whole cluster. K is a 
normalization 
constant. 

Finally, the conversion factor $\Gamma$ between the  total mass in the tail and 
the
number of stars
in the selected region (mask) of the cluster CMD is given by:

\begin{equation}
\Gamma=M_{tot}/N_{mask}=\frac{1-\alpha}{2-\alpha}\ 
\frac{m^{2-\alpha}_{up}-m^{2-\alpha}_{low}}
{M^{1-\alpha}_{max}-M^{1-\alpha}_{min}}
\end{equation}

$\Gamma$ is critically depending on the adopted value of $m_{low}$
and on the IMF slope $\alpha$  (see Scalo 1997
for an exhaustive discussion of the topic).

\tab{table_estimation} gives
the parameters  used  to compute the tidal tail mass in each object.
In the whole mass range a Salpeter 
slope ($\alpha=2.35$)
is adopted.

\begin{figure}
\resizebox{\hsize }{!}{\includegraphics{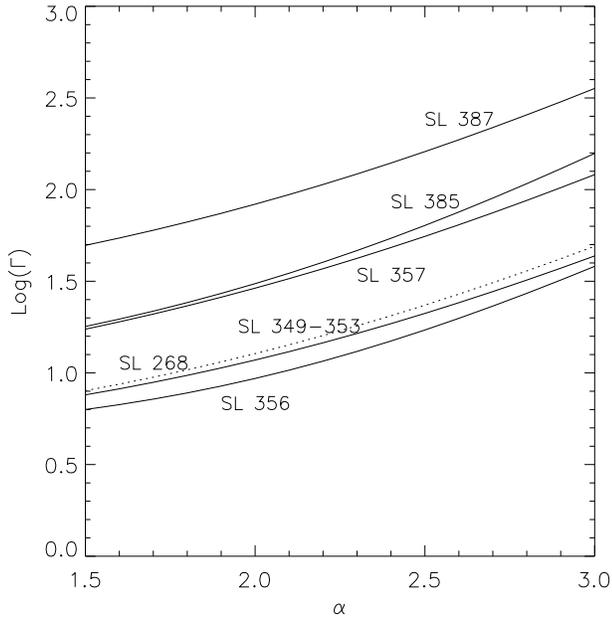}}
\caption{Gamma estimation for the studied
 clusters vs the slope of the mass function, 
relative
to the parameters of the CMD mask used to select the stars. The isolated cluster 
SL 268 is in 
dotted line. The data referring to NGC~1850 are not shown (see text for details) 
}
\label{fig_gamma_cluster}
\end{figure}

Several effects might influence our results:
 
(i) The clusters we have analyzed are affected by a severe crowding.
 This can result in a 
 underestimate  of the stellar density.
This effect becomes 
relevant mainly at fainter magnitudes and  towards the center
of the cluster. However, our star-counts are based on the
the brightest part of the CMD ($V > 20$), less 
affected 
by the crowding. 
From the usual experiments with artificial stars, we estimate that
the incompleteness correction, defined as the ratio of the number of
recovered stars to the total, is higher than 75-80\%
 for magnitudes brighter
than $V=20$, but becomes higher than 95\% for $V< 19$.\\
(ii) The $\Gamma$ conversion factor between the 
star-counts in the tail and the total  mass of the tail
is the most important source of uncertainty,  
 changing of 
an order 
of magnitude when the slope ranges  from 1.5 to 3.0.
 \fig{fig_gamma_cluster} 
shows the variation 
of $\Gamma$  in the observed clusters  at changing IMF slope.
NGC~1850  is not included, since the
  small field of view of NGC~1850 has prevent any reliable determination.
The result of \fig{fig_gamma_cluster} reflects
 the fact that the initial  mass function slope
in the LMC clusters is poorly constrained by the
observations. Additional difficulties arise when 
 mass segregation is present, as it is found
in the young LMC clusters SL 666 and NGC 2098 (Kontizas \etal 
1998). 
The $\Gamma$ factor is as well dependent on the minimum mass $M_{min}$ selected
by the CMD-masking. The $\Gamma$ estimate 
is less influenced by the value of the upper mass $M_{up}$.\\
(iii) To define the selected cluster region, we make use of
the diameters from Bica \etal (1996) which 
are not the tidal 
radii $r_t$. However, deriving the  tidal radius of both the components
of a pair  is 
quite difficult, because  
of the mutual interaction between the clusters.
In the case of SL 357, choosing a radius twice as larger results
 in  a mass determination 10\%
lower.
Nevertheless the correction  becomes more relevant  for  clusters having higher density  as shown in \tab{table_estimation}.\\ 
(iv) Pollution 
from overdensities not
related to the cluster cannot be avoided, as in the case of SL 385 with 
an overdensity 
in the N-E corner which is the cluster NGC 1926\,$\equiv$\,SL 403 (see 
\fig{fig_tail_cluster}). However, 
because of high S/N on the CMD mask, this effect is not expected to strongly 
bias
the  results.

Taking all these effects into account,
we estimate that the tidal mass indicated in \tab{table_estimation} is likely to 
be 
an upper limit to the 
real mass loss of the clusters.

\begin{table}
\begin{tabular}{llcccc}
\hline
\hline
Cluster	& $N_{tail}$ & $M_{min}$ & $M_{max}$ & $m_{up}$ & $M_{\rm {tail}}^\ddagger$/$M_{\rm{cl}}^{\S}$ \\
	& 	& (\msun)  & (\msun) & (\msun)  & ($10^3 \msun$)\\
\hline
SL 268   & 450(290) & 1.80	& 2.90	& 2.93 & 8.6 / 17\\
SL 349   & 450(190)$^\dagger$ & 	1.73 & 2.90 & 3.1 & 7.8\\
SL 353	 & 450(190)$^\dagger$ & 1.73 & 2.90 & 3.1 & 7.8\\
SL 356 	 & 130(50) & 1.90 & 6.64 & 6.64 & 1.8 / 26\\
SL 357 	 & 610(540)&  2.10	& 2.58	& 2.61 & 27.4\\
SL 385	 & 130(100) & 2.85 & 4.00 & 4.42 & 6.5\\	
SL 387	 & 230(180) & 2.35 & 2.54 & 2.83 & 29.9\\	
\hline
\end{tabular}
\\ $^\dagger$: no distinction between the two clusters SL 349 and SL 353.\\
$^\ddagger$: $M_{\rm {tail}}$ is the mass in the tails  given for a Salpeter slope~(2.35).\\ $^{\S}$: $M_{\rm {cl}}$ is the estimated total mass of the cluster.
\caption{$N_{tail}$ is the star count from the CMD mask in the tails, in 
parenthesis we give
the same estimate masking the cluster with a radius twice as larger.}
\label{table_estimation}
\end{table}

\section{Discussion}

In this section, 
we discuss the  results 
for each cluster pair separately. Hints about the most probable
formation scenario of binary clusters are given.

\subsection{NGC 1850}
\begin{figure*}
\resizebox{\hsize }{!}{\includegraphics{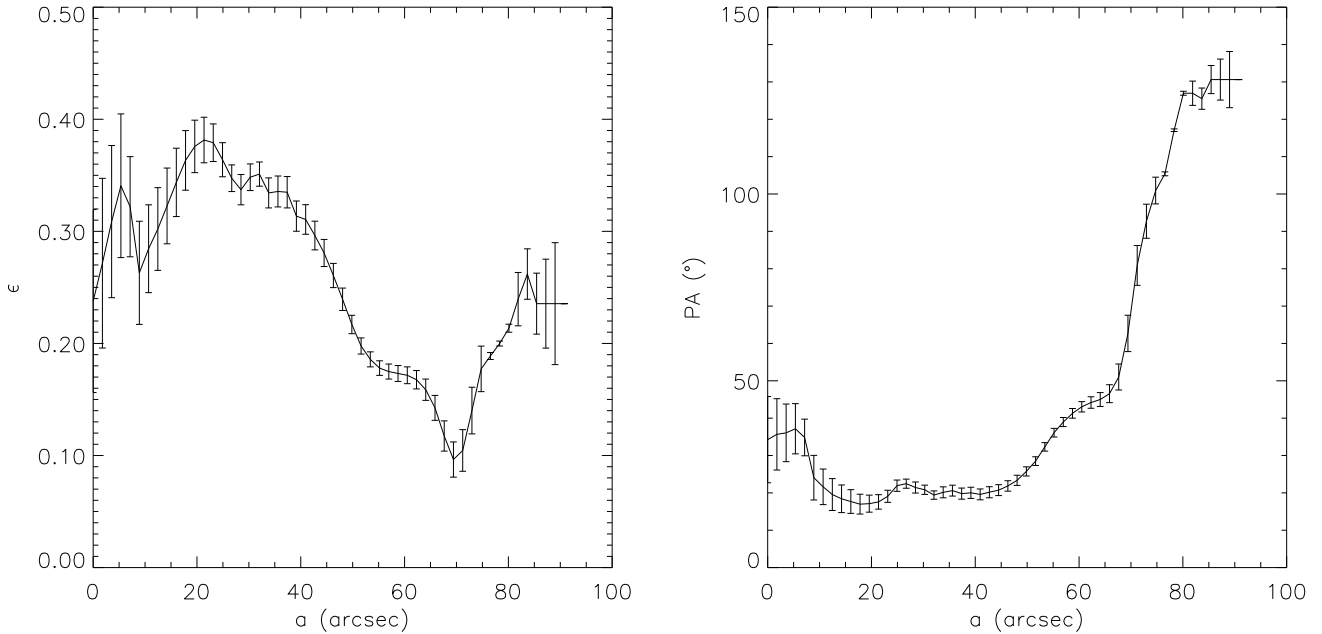}}
\caption{Ellipticity $\epsilon$ (left) and position angle PA (right) {\it vs.} 
semi-major axis for NGC 1850.}
\label{ellip1850}
\end{figure*}
NGC 1850 ($\equiv$ SL 261) is a massive cluster  which could be part of a binary or triple system 
(Fischer \etal 1993; Vallenari et al. 1994).
As already stated in previous sections,
 we did not perform  star-counts in the outermost halo of  
the cluster, because of the small field of view.We have performed an ellipticity 
and 
position angle analysis using {\tt ellipse} IRAF 
task that we show in \fig{ellip1850} to separate the core/halo structure
The surface density presents an elongated shape (see \fig{fig_tail_cluster})  
with an
ellipticity $\epsilon\sim0.25$
oriented in the direction NW-SE (PA$\sim$130\degree).
The  NW tail can be seen on the optical image as a 
concentration of stars. Contrary to the other clusters,
showing quite regular isophotes in the inner regions, NGC 1850 
exhibits very disturbed isophotes even at small radii
 which could be interpreted as
 due to the interaction with the 
other very close components.
The core structure is elongated perpendicularly to the outer  isophote 
(PA$\sim$20\degree). NGC 1850  is the only case in our sample exhibiting a 
high core ellipticity which could be attributed to strong tidal interactions 
in a triple system. 
 
 Fischer \etal (1993) noted the presence of two diffuse
clusters inside a radius of 6$\arcmin$ from NGC~1850 center.
 We tentatively interpret this structure
as due to stars 
 stripped 
 by
 the cluster
by gravitational ``harassment'' in this 
very crowded region at the border
 of the LMC bar.

 From a  ($V$, $B-V$) CMD,
Vallenari \etal (1994) found 3 components (called A, B, C)
located inside  about 1$\arcmin$
from the cluster center. They estimated the age of components
A and C to be about
$6\times 10^7$ yr, whereas the youngest component, B has an age
of $8 \times
10^6$ yr. The age difference between the components is marginally consistent
with the time scale expected on the basis of the model by Fujimoto \& Kumai 
(1997).

\subsection{SL 268}
SL~268 is an isolated cluster we have used to calibrate  the tidal effect of the 
LMC bar field on the 
clusters. Clearly the  interaction with the gravitational tidal field of the LMC 
has strong 
effect on the cluster evolution. The  NS tidal extension is 60 pc large.
In the southern 
part of it, the low density region is an artefact due to the presence of a 
bright star.
The tidal tail protruding towards NE 
accounts for a large 
part of the 8.6$\times 10^3$\msun{} present in the tails.

\begin{figure}
\resizebox{\hsize }{!}{\includegraphics{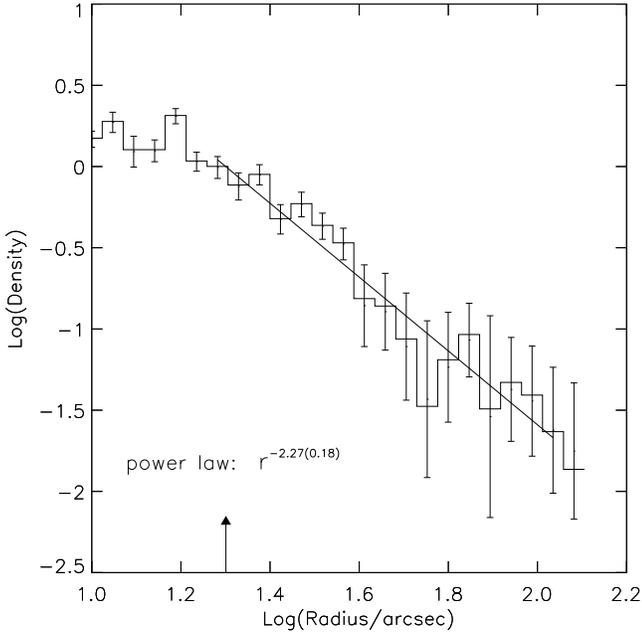}}
\caption{SL 268: surface density of stars selected from the CMD mask. The arrow 
is at the 
position of the cluster masking to count the halo and evaporated stars. The 
surface density is 
fitted by a power-law in the outskirts with a  slope of 2.27($\pm 0.18)$. To 
have an estimation 
of the surface density mass, the values have to be multiplied by the $\Gamma$ 
correction.}
\label{fig_sl268_profile}
\end{figure}

 A fit of the surface 
density in the halo of the cluster is performed, using a power law  function. 
The surface 
density, as shown in \fig{fig_sl268_profile}, is decreasing as 
$r^{-\gamma}$ with $\gamma$=2.27. This  is  consistent with the results by Elson 
\etal 
(1987, hereafter EFF) who found for a sample of LMC
clusters, typical values 
of $\gamma$ ranging from 2.2 to 3.2. 
However, we point out that, due to the tidal distortion of the isophotal 
contours,
 this one-dimensional  profile  is not a good 
representation of 
the spatial distribution which is far from being uniform with the
position angle. This 
profile, as found  by EFF, means that as much as 50\% of the total mass of 
the cluster 
could be in an unbound halo. 
We can estimate the  mass 
of SL 268 
to be about $1.7\times 10^4$\msun, following the procedure
outlined by EFF and taking the value of NGC 2004 in EFF which 
has a slope similar 
to the one of SL 268.
We remind that in the Galactic globular clusters
 the   tidal tail mass  represents only a few percents of the  
cluster mass (Combes \etal 1998;  Leon \etal 1998) where the tidal stripping
by the Galactic potential well is much stronger and efficient.

\subsection{Pair SL 349-SL 353}

Because of their similar CMD, SL 349 and SL 353 ($\equiv$ Hodge 1) cannot be separated, as 
discussed
in previous sections.
For this reason, no density profiles are presented.
 The total 
tidal tail mass of the pair is
$7.8\times 10^3$\msun{}. The tidal extension in the NS 
direction associated with 
SL 349 (see \fig{fig_tail_cluster}) is similar to the one of SL 268 and could be 
interpreted as due only to
the LMC tidal field.  The tail going towards the East  along 
the southern side of
SL 349 is difficult to be explained
 in that way. Even if a projection effect cannot be ruled out totally, it is tempting to associate it to the presence of  SL 353. A similar effect is noticed  in the case of SL 357 (see
following section).

According to Bhatia (1990), 
the survival time of binary system immersed in the LMC tidal 
field, is at maximum
of the order of $\sim 4\times 10^7$ yr. This pair, with an age of 
$5\times 10^8$ yr,
is an example of old surviving binary pair.
We refer to this as the ``overmerging'' problem.
This long lifetime might be explained by the scenario
 proposed by  Bica \etal (1992):
 star clusters form in groups (Star Cluster Groups, SCGs)  
inside Giant Molecular Clouds (GMCs) with  a high star formation efficiency. 
 The burst of 
cluster formations 
can be 
 enhanced during a  longer period than the one needed
to produce  a single cluster ($\sim 10^7$ years). This is in agreement
with Efremov \& Elmegreen (1998)
who find that the  star formation process
is hierarchical in time and space: small regions are expected to form stars on
short time scale (less than $10^6$ yr) , whereas large regions form stars over 
longer periods ($\ge 3 \times 10^7$ yr). 

If the clusters are formed as a part of SGC, it is not necessary 
that the pair  SL~349-SL~353
was  formed in
a bound binary system, but tidal capture at later time
could have been at work. In fact, inside such
a high density 
SCG  the rate of encounters must be higher than for isolated clusters
(see next section).

At the light of these works, we suggest that SL~349 and SL~353
are part of a SCG. Using the integrated photometry of LMC clusters by  
 Bica \etal (1996) we tentatively identify the objects
 forming this SCG.  We present the list in \tab{table_scg}.
The clusters are selected 
with the following criteria: cluster distance to 
``central'' (arbitrary) 
cluster less than 0.25\degree{} and a SWB type
(defined by Searle, Wilkinson \& Bagnuolo 1980)  differing at most of
 one unity from the 
SWB type of the
central cluster.

The star formation in the GMC 
could have been triggered by interaction between
LMC and SMC.
 Elmegreen \& 
Efremov (1997) have proposed that large-scale shocks can explain the formation 
of star clusters
in interacting galaxies. It is tempting to associate the younger SCGs to the 
interaction of
the LMC with the SMC which would have strongly enhanced the cluster formation a 
few $10^8$ years ago.

\subsection{Pair SL 356-SL 357} 

\begin{figure}
\resizebox{\hsize }{!}{\includegraphics{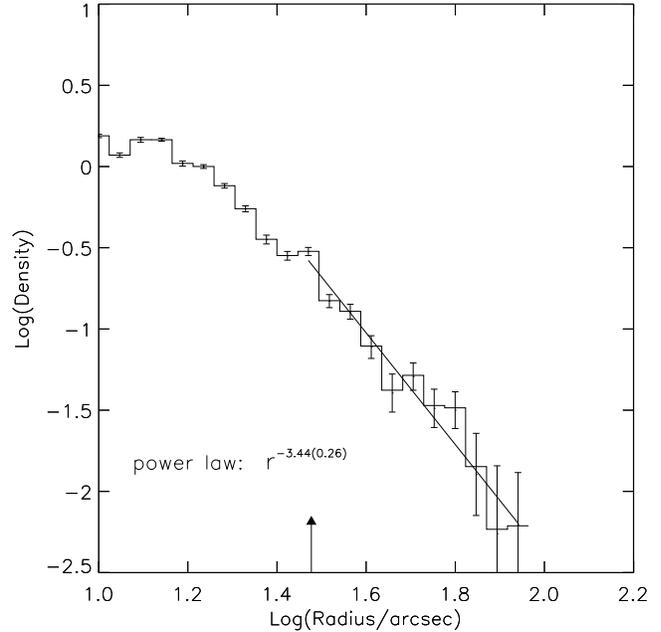}}
\caption{SL 356: surface density of stars selected from the CMD mask. The arrow 
is at the 
position of the cluster masked to  perform tidal tail star-counting. The surface 
density is 
fitted by a power-law in the outskirts with a  slope of $-3.44$ ($\pm 0.26$). 
To 
have an estimation 
of the surface density mass, the values have to be multiplied by the $\Gamma$ 
correction.}
\label{fig_sl356_profile}
\end{figure}

SL 356-SL 357 is a puzzling pair because of the large age difference between the 
two 
components, respectively 70 Myr for SL 356 and 600 Myr for SL 357. This age 
difference allows us to disentangle both the components on the basis of the
CMD properties.
  SL 356  ($\equiv$ NGC 1903) does not exhibit extended 
structures (see 
\fig{fig_tail_cluster}).  A southern tail is actually connected with the 
secondary 
component, already found by Vallenari \etal (1998).
In comparison with
 the other clusters (see \tab{table_estimation}), it has no  
 massive halo,
 which  is explained with its steep surface density profile in 
the outskirts 
(\fig{fig_sl356_profile}): a power law of slope $\gamma=3.44$  is consistent
with the data.

Following EFF method,  we derive for SL~356 the 
 total mass  of 
$2.6\times 10^4$\msun. 
SL 357 appears optically much weaker and more diffuse than SL 356 
(see \fig{fig_tail_cluster}). This object
 presents a very large tail which is 
obviously due to
its interaction with SL 356, as predicted by numerical simulations (de Oliveira et al. 1998).   

We point out that the age difference between
the two components is too large to be consistent with
the Fujimoto \& Kumai (1997) model of binary cluster formation.
Invoking a SCG scenario  for SL~356-SL~357 (see \tab{table_scg}),
the formation of this pair by tidal capture becomes more probable,
  because of an enhanced 
geometric cross section 
for the rate of encounters per cluster. Following Lee \etal (1995) we estimate a 
rate of encounters
$\frac{dN}{dt}\sim 10^8$ yr for a cluster crossing a SCG, which is compatible 
with the 
age of SL 357 and with the large separation of the two clusters (40 pc). 
The 
response to the 
interaction is different for the two clusters: it is plausible that
 the concentration of SL 356 was greater than the one of SL 357, as 
suggested by the steep
outer surface density. We have to point out that the mass of SL 356, estimated 
from its low halo mass
could be underestimated, while the estimation mass of the tail of SL 357 is 
likely overestimated, 
thus an unequal mass system cannot be ruled out as suggested by the Fig. 4 of 
de Oliveira \etal (1998) which is is very similar to this system.

\subsection{Pair SL 385-SL 387}

These two clusters present completely different spatial distribution in the 
halo: 
SL 387 exhibits
tidal tails typical of stellar systems in interaction (see e.g. Hibbard \& van 
Gorkom 1996 for interacting
galaxies; de Oliveira et al. 1998 for binary clusters)
 with a mass loss 5 times larger than  SL~385. 
 SL~385 presents a  modest extension   
closely resembling to 
  SL~268, and due at least partially, to the gravitational 
tidal field of the LMC.
The projected separation (10 pc) between the
pair components suggests that they will  merge rapidly 
in $\lesssim 10^7$ years (see 
Sugimoto \& Makino 1989; Bhatia 1990).

From the simulations of Sugimoto \& Makino (1989) 
for two  interacting
clusters, it appears that the stars lie in the trailing side of the orbiting 
binary: 
the orbital spin must  be an anti-clockwise orbital path in 
projection on the sky.
As in the case of the pair SL 349-SL 353, 
SL 385-SL 387 are both older than $10^8$ years: the 
overmerging problem
arises once more for such a close binary system.

The SCG formation scenario can 
represent a convincing explanation for the formation of this pair.
 The list of 
 clusters tentatively identified as belonging to this SCG together
with 
SL 385 and SL 387
is given in \tab{table_scg}. 
Finally, we point out that
due to the small separation of the pair, density profiles
cannot be used to derive the total masses of the clusters.

\begin{table}
\begin{tabular}{lcccccc}
\hline
\hline
\#1	 & \#2	& \#3	& \#4	& \#5  	&  \#6 & \#7\\
\hline
SL349 & SL353	& SL379 & SL390  & & &\\
SL356 & SL357 & SL344 & SL358 & SL361 & S\,Dor & SL373 \\
SL385 & SL387 & SL358 & SL403 & SL418 &\\
\hline
\end{tabular}
\caption{SCG memberships relative to the cluster observed (\#1). The criteria are the 
following:
projected distance to the cluster less than 0.25\degree{} and with SWB type 
different at most
by one unity (from Bica \etal 1996).}
\label{table_scg}
\end{table}

\section{Conclusion}

We  study the tidal tail extensions of 8 star clusters in the LMC bar, 
using
 star-counts on the CMDs. Seven of these clusters are suspected to belong
to binary or multiple systems, while one (SL~268) is isolated.
The presence of tidal tails typical of stellar systems in interaction
confirms that all the  pairs are physically connected, as already found for other binary clusters 
by Kontizas et al. (1993): it appears that in their cluster sample, the ages of the binaries are 
less than a few 10$^7$ yrs and a coeval evolution before the merging cannot be ruled out. 
Nevertheless we point out that their objects are located in a cluster-rich environment which can 
be related to a SCG.

 The isolated cluster SL 268 have 
about 50\% of its mass in
its outskirts, as found already for other clusters by EFF. The halo 
density distribution of this cluster appears to be
strongly shaped by the LMC tidal field. The luminosity profiles
of  SL 268 and SL 356, at large 
radii
 fall off as $r^{-\gamma}$,  with $\gamma$ equal to $2.27$ 
and $3.44$, respectively, in good agreement with  the
values found by EFF.

All the observed pairs are older than
the survival time of binary clusters
estimated by theoretical models (less than 
$4\times 10^7$ years, see Bhatia 1990; Sugimoto \& Makino 1989).
We refer to this  as the 
``overmerging problem''.
The star cluster group (SCG) scenario invoking 
the formation of these clusters
as  part of a large  group  during the 
last interaction with 
the SMC, can help in solving this problem. We point out that a recent study of binary clusters n 
the SMC discloses the same ``overmerging problem'' (Dieball \& Grebel, 1998).
We suggest that this scenario formation in SCG induced by tidal interaction with the SMC is different
with the quiescent one in the Milky Way: the binary cluster frequency appears  to be much lower, despite
the opposite claim of Subramaniam \etal (1995). 
In the SCGs  binary clusters could have formed 
 with a delay 
compatible with the current age
found in the studied pairs.
 This scenario, as well, could explain the peculiar pair SL 
356-SL 357 which would 
have form from a tidal capture of SL 357 by the more massive cluster
SL 356, thanks to a 
cross section increased in the
SCG. 
 We tentatively give in 
 the memberships 
of the different SCGs selected from the
survey of Bica \etal (1996).
A dynamical study of the 
SCGs would be of
great interest. A more complete identification of SCG and a dynamical analysis 
of star clusters
embedded in the LMC field will be presented in forthcoming papers.
\begin{acknowledgements}
We thank G. Meylan  for his helpful remarks and V. Charmandaris for his {\sf 
xIRAF} scripts. We also gratefully acknowledge useful comments from the referee M.~Kontizas.
\end{acknowledgements}

\end{document}